\documentclass[aps,pra,twocolumn,showpacs,showkeys,amsmath,superscriptaddress,floatfix]{revtex4}

\usepackage{amsmath}
\usepackage{makeidx}
\usepackage{amsfonts}
\usepackage[utf8]{inputenc}
\usepackage[usenames,dvipsnames]{pstricks}
\usepackage{subfigure}
\usepackage{epsfig}
\usepackage{pst-grad} 
\usepackage{pst-plot} 
\usepackage[colorlinks,hyperindex]{hyperref}
\hypersetup
{
	colorlinks,%
	citecolor=black,%
	linkcolor=black,%
	urlcolor=black,%
}

\renewcommand{\vec}[1]{\mathbf{#1}}

\begin{document}
	
	\title{Dynamical vanishing of the order parameter in a confined Bardeen-Cooper-Schrieffer Fermi gas after an interaction quench}
	
	\author{S.~Hannibal}
	\affiliation{Institut f\"ur Festk\"orpertheorie, Westf\"alische
		Wilhelms-Universit\"at M\"unster, 48149 M\"unster,
		Germany}
	 
	\author{P.~Kettmann}
	\affiliation{Institut f\"ur Festk\"orpertheorie, Westf\"alische
		Wilhelms-Universit\"at M\"unster, 48149 M\"unster,
		Germany}
	
	\author{M.~D.~Croitoru}
	\affiliation{Theoretische Physik III, Universit\"at Bayreuth, 95440
		Bayreuth,
		Germany}
	
	
	\author{V.~M.~Axt}
	\affiliation{Theoretische Physik III, Universit\"at Bayreuth, 95440
		Bayreuth,
		Germany}
	
	\author{T.~Kuhn}
	\affiliation{Institut f\"ur Festk\"orpertheorie, Westf\"alische
		Wilhelms-Universit\"at M\"unster, 48149 M\"unster,
		Germany}
	
	\date{\today}
	
	\begin{abstract}
	We present a numerical study of the Higgs mode in an ultracold confined Fermi gas after an interaction quench and find a dynamical vanishing of the superfluid order parameter. Our calculations are done within a microscopic density-matrix approach in the Bogoliubov-de Gennes framework which takes the three-dimensional cigar-shaped confinement explicitly into account. In this framework, we study the amplitude mode of the order parameter after interaction quenches starting on the BCS side of the BEC-BCS crossover close to the transition and ending in the BCS regime. We demonstrate the emergence of a dynamically vanishing superfluid order parameter in the spatiotemporal dynamics in a three-dimensional trap. Further, we show that the signal averaged over the whole trap mirrors the spatiotemporal behavior and allows us to systematically study the effects of the system size and aspect ratio on the observed dynamics. Our analysis enables us to connect the confinement-induced modifications of the dynamics to the pairing properties of the system. Finally, we demonstrate that the signature of the Higgs mode is contained in the dynamical signal of the condensate fraction, which, therefore, might provide a new experimental access to the nonadiabatic regime of the Higgs mode.
	\end{abstract}
	
	
	
	\maketitle
	
	\section{Introduction}\label{sec:Introduction}
	Due to a remarkable control over many relevant system parameters which, in most cases, can be tuned at will, ultracold quantum gases are an ideal testbed for many-body theories and for concepts known from solid-state theory. These include, e.g., lattice symmetries \cite{Bloch2008Many}, topological properties \cite{Goldman2016Topological}, spin-orbit coupling \cite{Galitski2013Spin, Wu2016Realization}, and non-homogeneous superconductivity \cite{Liao2010Spin}. For these reasons ultracold Fermi gases have received great attention both from an experimental and from a theoretical point of view \cite{Bloch2008Many,Giorgini2008Theory}.
	
	Using a Feshbach resonance \cite{Chin2010Feshbach} to adjust the interaction strength a Bardeen-Cooper-Schrieffer (BCS) superfluid state of Cooper pairs has been achieved for small attractive interactions, whereas in the limit of large interactions between the fermions repulsive bosonic dimers form and a Bose Einstein condensate (BEC) of the latter has been observed \cite{Regal2004Observation}. Those two limiting superfluid states are connected by the BCS-BEC crossover featuring strong interactions and unitary properties \cite{Zwerger2011BCS}.
	
	The implementation of ultracold Fermi gases always requires an external confinement, e.g., optical traps. The control of this external confinement on the one hand provides means for the outstanding control over the system \cite{Bloch2008Many,Grimm2000Optical}. On the other hand, this implies a finite system size which can have a major impact on the physics of a superfluid system. The pairing properties strongly depend on the structure of the energy spectrum of the system and, hence, on the dimensionality of the system. Theoretical studies predict an atypical BCS-BEC crossover due to confinement effects  \cite{Shanenko2012Atypical}.
	
	Recently, efforts have been devoted to the non-equilibrium properties of the superfluid state \cite{Polkovnikov2011Colloquium:,Yin2013Quench,Yin2016Quench}. Due to the spontaneously broken U(1) symmetry two fundamental modes of the complex order parameter evolve: the (massive Higgs) amplitude mode and the (massless Goldstone) phase mode. The observation of the collective Higgs excitation in a system with spontaneous symmetry breaking (SSB) is of fundamental importance to gain a better understanding of the physical system at hand, as was recently demonstrated in the case of the standard model of particle physics \cite{Chatrchyan2012Observation}. 
	
	The first evidence of the Higgs mode was reported by Raman scattering in a superconducting charge density wave compound \cite{Sooryakumar1980Raman,Littlewood1981Gauge}. Furthermore, in ultracold gases the Higgs mode has been observed in the spectral response of a 2D optical lattice excited by an amplitude modulation of the lattice \cite{Endres2012Higgs}. A time-resolved observation of the nonadiabatic regime of the Higgs mode \cite{Papenkort2007Coherent,Papenkort2008Coherent}, as has been achieved in a superconducting NbN film \cite{Matsunaga2014Light,Matsunaga2013Higgsa,Matsunaga2012Nonequilibrium}, is still pending for ultracold gases. In order to reach the nonadiabatic regime the challenging implementation of an interaction quench has been proposed to be done either by an RF flip of one species of the condensate (cf. \cite{Froehlich2011Radio}), which leads, in the vicinity of a Feshbach resonance, to the desired almost instantaneous change of the scattering length. Alternatively the quench could be implemented by an optical control of the interaction in quantum gases \cite{Clark2015Quantum}. However, there are several theoretical studies addressing an ultracold Fermi gas after an interaction quench \cite{Yuzbashyan2015Quantum,Yuzbashyan2006Relaxationa,Yuzbashyan2006Dynamical,Yuzbashyan2005Nonequilibrium,Barankov2006Synchronization,Bruun1999BCS,Bruun2014Long,Scott2012Rapid,Hannibal2015Quench}. In the case of a homogeneous system three dynamical phases depending on the excitation condition have been predicted analytically  \cite{Yuzbashyan2015Quantum}. They include persistent oscillations of the BCS gap (``phase III''), damped oscillations with inverse square root decay (``phase II'') and a dynamical vanishing of the order parameter (``phase I''). Previous studies in confined systems focused on the damped oscillations and showed that for superconducting BCS nanowires the decay exponent is changed from $-1/2$ to $-3/4$ due to quantum size resonances \cite{Zachmann2013Ultrafast}, while for even tighter confined superconducting nanorods the decaying oscillation becomes irregular \cite{Kettmann2017Spectral}. In a BCS Fermi gas confined in a box with periodic boundary conditions in two dimensions and a harmonic confinement in the third dimension the inverse square-root decay has been confirmed \cite{Scott2012Rapid} and for a three-dimensional harmonic confinement an additional fragmentation of the damped oscillations has been predicted \cite{Hannibal2015Quench}.
	
	In this paper we show that the dynamical vanishing of the superfluid order parameter, i.e., phase I, also emerges in a Fermi gas confined in a three dimensional harmonic trapping potential after an interaction quench. The emergence of phase I has not only been shown in ultracold Fermi gases but also in BCS superconductors \cite{Papenkort2009Nonequilibrium,Chou2017Twisting}. Here, we analyze the transition to a dynamically vanishing order parameter in dependence of the confinement properties. To this end, we present a numerical study based on the previously published microscopic density matrix approach within the Bogoliubov-de Gennes framework \cite{Hannibal2015Quench}. This allows for a full microscopic and coherent quantum mechanical treatment of the system and provides access to the spatiotemporal dynamics of the order parameter as well as to the condensate fraction which is a possible experimental candidate to carry the signature of the Higgs mode.
	
	
	This paper is organized as follows. In Sec. \ref{sec:model} we will give a short summary of the used theoretical model, while Sec. \ref{sec:results} is devoted to the results of our numerical calculations. We start by discussing the spatiotemporal dynamics of the order parameter in Sec.~\ref{sec:spatiotemporal}. In the next step, we will give a detailed analysis of the dynamical vanishing in a cigar-shaped trap in Secs. \ref{sec:vanish} and \ref{sec:aspect_ratio} based on the spatially averaged order parameter. In Secs.~\ref{sec:scaling} and \ref{sec:large} we will provide extrapolations to typical experimental particle numbers. Finally, in Sec. \ref{sec:condensate} we present the dynamics of the condensate fraction after an interaction quench. In Sec. \ref{sec:conclusions} we will briefly summarize our findings.

	\section{Theoretical Model}\label{sec:model}
	
	 We employ the Bogoliubov-de Gennes (BdG) formalism \cite{Hannibal2015Quench, Datta1999Can, DeGennes1989Superconductivity} to calculate the dynamics of the superfluid order parameter $\Delta(\vec{r},t)$ of an ultracold Fermi gas confined in an axially symmetric harmonic trapping potential after an interaction quench. In this section, we provide a short summary of the most essential aspects of the used formalism and the relevant system parameters. For a comprehensive discussion of the formalism we refer the reader to our previous work \cite{Hannibal2015Quench}.
	 
	 We consider the gas to be composed of fermionic $^6$Li atoms in two different internal spin states (labeled by $\uparrow$ and $\downarrow$) with equal particle numbers $N_\uparrow = N_\downarrow = N /2$. The atoms in the two spin states interact via a contact interaction with $V_\text{contact}=(4 \pi \hbar^2 a / m )\, \delta(\vec{r}_1 -\vec{r_2}) = g\, \delta(\vec{r}_1 -\vec{r_2})$, where $m$ is the mass of a $^6$Li atom and $a < 0$ is the scattering length. Hence we only consider systems on the BCS side of the BEC-BCS crossover.
	 
	 We use a cigar-shaped harmonic confinement potential given by $ V_\text{conf} = \frac{1}{2}m \omega_\perp^2 (x^2+y^2) + \frac{1}{2}m \omega_\parallel^2 z^2$, where $\omega_\perp$ $(\omega_\parallel)$ is the radial (longitudinal) confinement frequency, respectively. We denote the eigenfunctions of the confinement potential by $\left<\hat{r}|k\right> = \phi_k(\vec{r})$ and the equidistant single-particle energies of the confinement potential are labeled by $\xi_k$, where we use the tuple $k = (k_x,k_y,k_z)$ with $k_i \in \{0,1,2,..\}$. In this paper, we consider elongated traps characterized by an aspect ratio of the cloud $r = \omega_\perp / \omega_\parallel \gg 1$. In such a geometry one-dimensional subbands form resulting in quantum size oscillations in the case of small particle numbers \cite{Shanenko2012Atypical, Hannibal2015Quench}. In order to characterize the number and positions of these subbands it is instructive to introduce the subband parameter $s = E_F / (\hbar \omega_\perp)$ where the Fermi energy $E_F$ is given by the chemical potential in the non-interacting case. That is, for a system with $s=j$, where $j \in \mathbb{N}$, the minimum of subband $j$ is located at the Fermi energy. The choice $s = j + 0.5$ implies that the physics of the system are not dominated by a single band and the effects of the quantum size resonances are negligible also in small systems (cf. \cite{Hannibal2015Quench}).

	 
	 In order to describe the system, we use the introduced contact interaction and confinement potential and we obtain the BdG Hamiltonian with the BCS-like mean-field approximation 
	 \begin{align}\label{eq:Hamiltonian}
	 H_\text{BdG} = \sum_\sigma &\int d^3r\,  \Psi_\sigma^\dagger (\vec{r})  H_0 \Psi_\sigma(\vec{r}) \notag \\  + &\int d^3r \, \Delta^*(\vec{r}) \Psi_\downarrow(\vec{r})  \Psi_\uparrow(\vec{r}) + h.c. \, ,
	 \end{align}
	 where $H_0$ is the single-particle Hamiltonian. By doing so we introduce the superfluid spatially dependent order parameter 
	 \begin{equation} \label{eq:order}
		\Delta(\vec{r}) = g \left< \Psi_\downarrow(\vec{r})\Psi_\uparrow(\vec{r})\right>.
	 \end{equation}
	  Here, $\Psi_{\sigma}(\vec{r})$ is the field operator annihilating an atom with spin $\sigma$ at position $\vec{r}$. The corresponding eigenvalue problem to the Hamiltonian in Eq.~\eqref{eq:Hamiltonian} is the BdG equation which reads
	  \begin{equation}\label{eq:Bogoliubov-Gl}
	  \begin{pmatrix} H_0 & \Delta(\vec{r}) \\ \Delta^*(\vec{r}) & -H_0^* \end{pmatrix}
	  \left(\begin{array}{c} u_K(\vec{r}) \\ v_K(\vec{r}) \end{array} \right)
	  = E_K \left( \begin{array}{c} u_K(\vec{r}) \\ v_K(\vec{r}) \end{array} \right).
	  \end{equation}
	  
	  The solution of this equation yields two branches (labeled by $K\rightarrow ka$ and $K \rightarrow kb$) for the eigenenergies and eigenfunctions which result from the two spin species. However, the eigenenergies and eigenfunctions can be expressed by one another \cite{Datta1999Can}. Therefore, we drop the index \textit{a/b} wherever possible and imply thereby the use of the states and energies corresponding to branch \textit{a}.
	  
	  Since the BdG equation has the form of a one-particle Schrödinger equation we use the BdG wave functions to introduce Bogoliubov's quasiparticles which diagonalize $H_\text{BdG}$. The transformation reads
	  \begin{align}\label{eq:Bogolon}
	  \gamma^\dagger_{ka} = \int u_k(\vec{r}) \Psi_\uparrow^\dagger(\vec{r}) + v_k(\vec{r}) \Psi_\downarrow(\vec{r})\, d^3 r \notag \\ 
	  \gamma^\dagger_{kb} = \int u_k(\vec{r}) \Psi_\downarrow^\dagger(\vec{r}) - v_k(\vec{r}) \Psi_\uparrow(\vec{r})\, d^3 r.
	  \end{align}
	  
	  In order to obtain the solution of the BdG equation we make use of Anderson's approximation \cite{Anderson1959Theory}, i.e., we assume the BdG wave functions to be proportional to the bare atomic wave functions with $u_k(\vec{r}) = u_k \phi_k(\vec{r})$ and $v_k(\vec{r}) = v_k \phi_k(\vec{r})$.  This yields the BdG eigenenergies
	 \begin{equation}\label{eq:Ek}
	   E_k = \sqrt{\left(\xi_k-\mu\right)^2+\Delta_k^2}
	 \end{equation}
	 and BdG coefficients
	 \begin{equation}
	  u_k = \sqrt{\frac{1}{2} \left(1+ \frac{\xi_k-\mu}{E_k}\right)} \quad  v_k = \sqrt{\frac{1}{2} \left(1 - \frac{\xi_k-\mu}{E_k}\right)}
	 \end{equation}
	 where $\Delta_k = \left<k|\Delta(\vec{r})|k\right>$.
	 Exploiting this solution and substituting Bogoliubov's quasiparticles in the definition of the order parameter in Eq.~\eqref{eq:order} we obtain the well-known self-consistency equations for the order parameter and the chemical potential $\mu$ in the ground state of the system at $T=0\,$K. They read
	 \begin{align}
	  \Delta_k &= - \frac{1}{2} \sum_{k'} V_{kk'} \frac{\Delta_{k'}}{E_{k'}} \chi_{k'} \label{eq:gap} \\  
	  N &= 2 \sum_k \left(1 - \frac{\xi_k - \mu}{E_k}\right), \label{eq:particle}
	 \end{align}
	  where we have used $V_{kk'} = g \int d^3r \left|\phi_k(\vec{r})\right|^2 \left|\phi_{k'}(\vec{r})\right|^2 $. The self-consistent solution $\Delta_k$ of these equations characterizes the pairing properties in the ground state of the system, where the pairing takes place predominantly in an interval around the chemical potential as known from bulk BCS theory. The factor $\chi_{k'} = (1 - E_{k'} / \xi_{k'} ) $ in Eq.~\eqref{eq:gap} regularizes the well-known ultraviolet divergence of the contact interaction \cite{Bruun1999BCS,Bruun2002Cooper,Shanenko2012Atypical}. This regularization is introduced into all terms which result from summations over states $k$. Additionally, the numerically necessary cutoff is chosen such that all physical effects are qualitatively correctly accounted for.
	  
	  In order to excite the nonadiabatic dynamics of the system we implement an interaction quench, i.e., an instantaneous change of the scattering length $a$, which leaves the particle density $n(\vec{r}) = n_\uparrow (\vec{r}) + n_\downarrow(\vec{r}) = 2 \left<\Psi_\uparrow^\dagger \Psi_\uparrow^{\phantom{\dagger}}\right>$ unchanged. In doing so, only occupations $x_{kl} = \left<\gamma_{ka}^\dagger \gamma_{la}^{\phantom{\dagger}}\right> = \left<\gamma_{kb}^\dagger \gamma_{lb}^{\phantom{\dagger}}\right> $ and coherences $y_{kl} = \left<\gamma_{ka}^{\dagger}\gamma_{lb}^{{\dagger}}\right> = \left<\gamma_{lb}^{\phantom{\dagger}} \gamma_{ka}^{\phantom{\dagger}}\right>^*$ which are diagonal in terms of the quasiparticles, i.e., $x_{kl}=x_k\,\delta_{kl}$ and $y_{kl} = y_k\,\delta_{kl}$, are excited.
	  
	  In the next step, we utilize Heisenberg's equations of motion in order to obtain equations of motion for the occupations $x_k$ and coherences $y_k$ of the density matrix. For our dynamical calculations we choose a time-independent basis, i.e., the basis given by the final system ($u_m$,$v_m$). This leads to a time-dependent Hamiltonian and the appearance of terms specifying the deviation between the dynamical gap $\Delta$ and the ground-state value $\Delta_{GS}$. The obtained equations of motion read
	  \begin{align}
	  i \hbar \frac{d}{dt} x_k &= a_k y_k^* - a_k^* y_k \label{eq:Bwg-Gl-normal}\\
	  i \hbar \frac{d}{dt} y_k & = -2\, E_k^{\text{(ren)}} y_k + a_k \left( 1 - 2 x_k \right) \label{eq:Bwg-Gl-anomal},
	  \end{align}
	  where 
	  \begin{align}
	  E_k^{\text{(ren)}} &= E_k + 2 \, u_k v_k \rm{Re}\left[(\Delta-\Delta_{GS})_k\right] \\
	  a_k &= v_k^2(\Delta-\Delta_{GS})_k - u_k^2(\Delta-\Delta_{GS})_k^*
	  \end{align}
	  and the deviation of the current gap is given by 
	  \begin{equation}
	   (\Delta - \Delta_{GS})_{k} = -\sum_{l} \left[ 2 v_l u_l x_l	+ u_l^2 y_l^*- v_l^2 y_l\right] V_{kl} \chi_l.
	  \end{equation}
	  Here, $\chi_l$ regularizes this term in analogy to Eq.~\eqref{eq:gap}. We solve these equations of motion numerically with the Runge-Kutta-Fehlberg method which provides good numerical stability for all investigated sets of parameters.
	
	  Finally, we invert Bogoliubov's transformation and we arrive at the space- and time-dependent order parameter in terms of the quasiparticles which reads
	  \begin{align}\label{eq:Delta_Bogolon}
	      \Delta(\vec{r},t) =  g  \sum_{k} \, & 2\, v_k(\vec{r}) u_k(\vec{r})\, \left( x_k(t) - \frac{1}{2}\right) \notag \\ &+ \, u_k^2(\vec{r})\, y_k^*(t) - v_k^2(\vec{r})\, y_k(t).
	  \end{align}

	  \section{Results}\label{sec:results}
	  In this section, we will first discuss the full spatiotemporal dynamics of the modulus of the order parameter $\left|\Delta(\vec{r},t)\right|$ and in the next step we turn to the temporal evolution of the modulus of the spatially averaged order parameter given by
	  \begin{equation}
	   \overline{\Delta}(t) = \left|\frac{1}{V} \int d^3r \Delta(r,t)\right|,
	  \end{equation}
	  where the volume $V = (\hbar/m)^{3/2} \omega_\perp^{-1} \omega_\parallel^{-1/2} = l_\perp^2 l_\parallel$ is determined by the harmonic trapping frequencies. This quantity will enable us to carry out a systematic study of the obtained dynamics in dependence on the trap parameters.
	  
	  In this paper, we investigate interaction quenches for which the initial ground state, characterized by $1/(k_Fa_i)$ and the ground-state gap $\overline{\Delta}_i$, is situated in the crossover regime of the BEC-BCS crossover and the final ground state [$1/(k_Fa_f),\,\, \overline{\Delta}_f$], is situated deep in the BCS regime. Here, we assume the well-known dispersion relation of the homogeneous Fermi gas in order to obtain the Fermi wave vector $k_F$. Throughout the paper we choose $1/(k_F a_f) = -1.45$ and characterize the strength of the quench by the initial coupling parameter $1/(k_Fa_i)$. We note that it would also be possible to characterize the quench strength by $\overline{\Delta}_f/\overline{\Delta}_i$ in analogy to \cite{Yuzbashyan2015Quantum}. However, with the introduced regularization $\overline{\Delta}_f/\overline{\Delta}_i$ converges much slower than $\Delta_k$ when increasing the numerical cutoff. Since $\Delta_k$ determines the main properties of the system in our formalism it is sufficient to choose the numerical cutoff according to $\Delta_k$, which substantially reduces the numerical effort. Bearing this in mind, we characterize the quench strength by $1/(k_Fa_i)$, which is independent of the numerical cutoff. In the case of a homogeneous system a study of similar quenches has been done before and our quenches here are connected to the emergence of phase I \cite{Yuzbashyan2015Quantum, Yuzbashyan2006Dynamical}. 
	  
	  \subsection{Spatiotemporal dynamics}\label{sec:spatiotemporal}
	  
	  In this section, we will investigate the spatiotemporal dynamics of the order parameter. To this end, we keep here, and in Secs.~\ref{sec:vanish} and \ref{sec:aspect_ratio}, the subband parameter $s = 3.5$ fixed. 
	  Furthermore, we choose $\omega_\parallel = 2\pi \cdot 120\,$Hz, which then determines together with a given aspect ratio $r$ and subband parameter $s$ the transverse trapping frequency $\omega_\perp$ and the particle number $N$.
	  
	  \begin{figure}[t]
		\includegraphics[width=1\columnwidth]{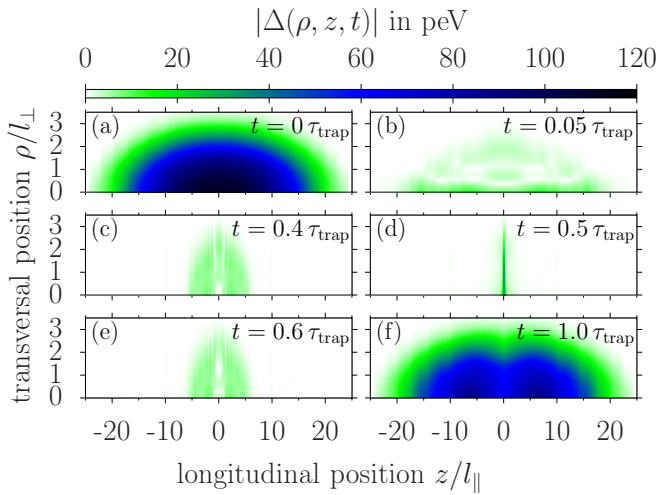}
		\caption{Spatiotemporal dynamics of the modulus of the order parameter after a sudden change of the scattering length from $1/(k_Fa_i) = -0.4$ to $1/(k_Fa_f) = -1.45$ and an aspect ratio of $r=50$, where $\tau_\text{trap} = h/\delta E = 1/(2 f_\parallel)$ is the characteristic time scale of the trap \cite{Hannibal2015Quench}.}
		\label{fig:spatiotemporal}
	  \end{figure}
	  
	  In Fig.~\ref{fig:spatiotemporal} we show the spatially dependent modulus of the order parameter $\left|\Delta(\vec{r},t)\right|$ at six different times after an excitation by an interaction quench. Figure~\ref{fig:spatiotemporal}(a) shows the initial spatial distribution of the gap at $t=0$ after the instantaneous quench, i.e., the excitations introduced by the quench are already included. We find a symmetrical distribution which features a maximum in the center of the trap. After $t=0.05\tau_\text{trap}$, we obtain a rapid decay of the order parameter of approximately one order of magnitude which affects the whole trap (cf. Fig.~\ref{fig:spatiotemporal}(b)). In addition, we find that the relative suppression of the order parameter with respect to the ground-state value of the final system is strongest in the center of the trap. Subsequently, we find that the remaining condensate moves towards the center of the trap where the distribution of the order parameter becomes very narrow with a large amplitude at $t= 0.5\tau_\text{trap}$ (cf. Fig.~\ref{fig:spatiotemporal}(d)). Comparing Figs.~\ref{fig:spatiotemporal}(c) and (e), which are almost identical, we see that this behavior occurs oscillatory and we find that the frequency is given by $f_\text{collapse}= 2 f_\parallel$. Finally, in Fig.~\ref{fig:spatiotemporal}(f) we see a revival of the order parameter at $t = \tau_\text{trap}$, which shows two maxima located symmetric to the center of the trap.
	  
	  Overall, from Fig.~\ref{fig:spatiotemporal} we extract two effects. The first one is characterized by the rapid initial decay of the order parameter and the revival after $\tau_\text{trap}$. The initial distribution of $\left|\Delta(\vec{r},t=0)\right|$ is set by the ground-state order parameter of the initial system while at the time of the revival we see that the order parameter shows two maxima which result from --as our data shows-- the spatial profile of the order parameter in the final system. The second effect is an oscillation taking place in the remaining superfluid after the initial decay. We see from our data that the suppression of the order parameter with respect to the ground-state value is strongest in the center of the trap. This causes the superfluid to start oscillating in the harmonic trap by moving towards the center. An analysis of the complex phase of the order parameter $\Phi(\vec{r},t)$ shows that this oscillation fulfills the general relation between the superfluid velocity and the gradient of the phase, i.e., $v_s = \hbar / m \, \nabla \Phi(\vec{r},t)$. This implies that the order parameter obeys a continuity equation and, hence, this oscillation will not be visible in the spatially averaged order parameter, as we will see in the next section.
	  
	  Before we proceed, we point out that during the dynamics the average occupation number of the fermionic atoms stays smeared out around the chemical potential and the coherences $y_k$ of the density matrix stay finite, indicating the existence of pair correlations at all times. In equilibrium these properties are directly linked to a superfluid behavior of the system. In a dynamical situation these properties are typically related to a superfluid behavior of the system even in the case of a vanishing order parameter as was also pointed out in Ref. \cite{Yuzbashyan2006Dynamical}.
	  
	  In the following, we will systematically analyze the dynamical vanishing of the order parameter in a harmonic trap. To this end, we will discuss the modulus of the order parameter averaged over the trap. First, we will show that this is a suitable quantity to characterize the dynamical vanishing. Further, we will investigate its dependence on the aspect ratio $r$ of the trap for small systems with fixed parameter $s$. In the next step we will show how the results scale for a larger system and, finally, we will present the dynamics of the condensate fraction for larger systems after the same type of quench.
	  
	  \subsection{Dynamical vanishing in the spatially averaged order parameter}\label{sec:vanish}
	  
	  \begin{figure}[t]
		\includegraphics[width=1\columnwidth]{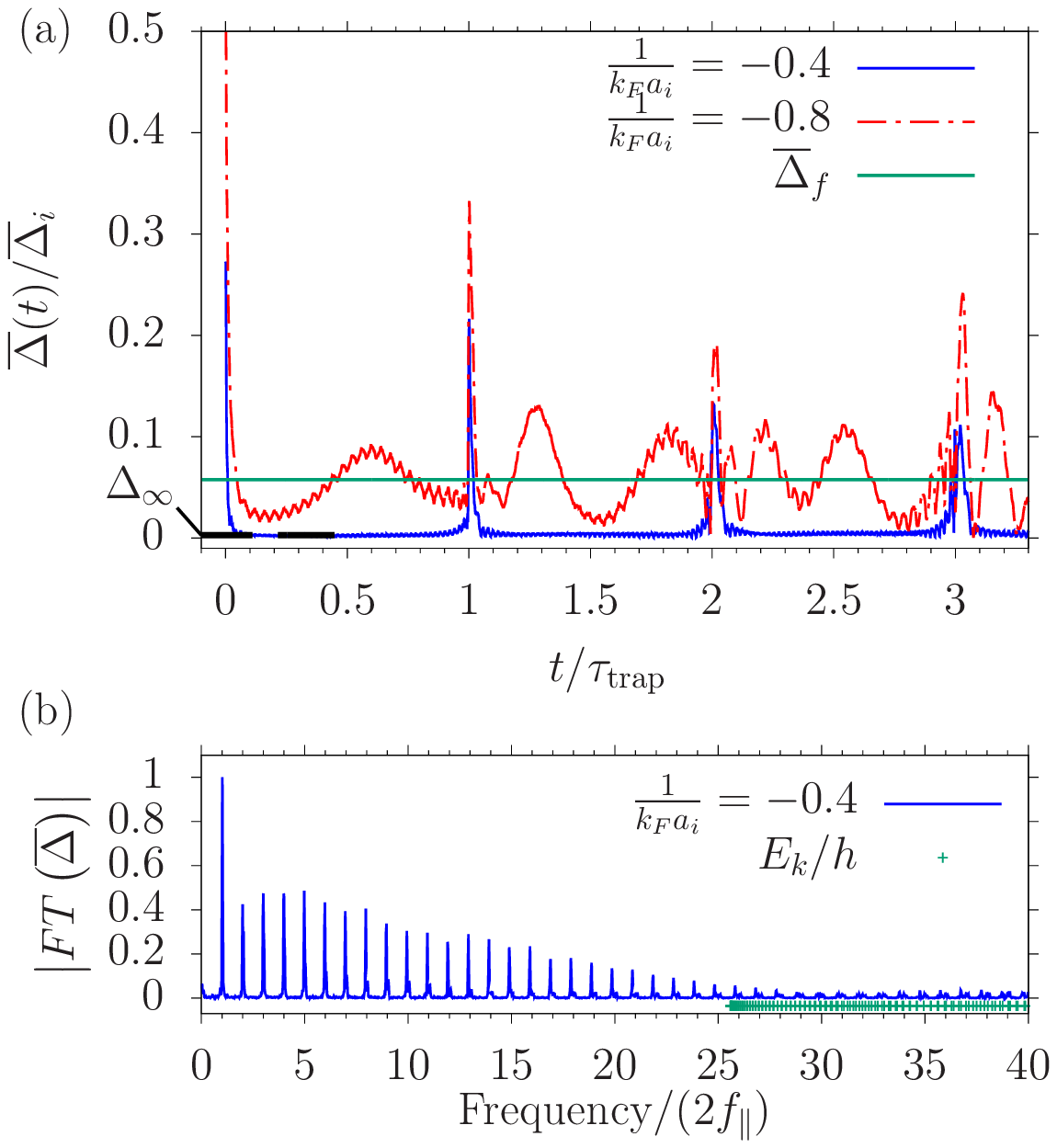}
		\caption{(a) Dynamics of the spatially averaged gap after a sudden change of the
			scattering length from $1/(k_Fa_i) = -0.4$ (blue solid line) and $1/(k_Fa_i) = -0.8$ (red dash-dotted line) to $1/(k_Fa_f) = -1.45$ and an aspect ratio of $r=50$. The gap is normalized to the spatially averaged ground-state gap of the system before the quench $\overline{\Delta}_i$. $\overline{\Delta}_f$ is the spatially averaged ground-state gap of the final system and $\Delta_\infty$ marks the height of the plateau for $1/(k_Fa_i) = -0.4$; (b) Fourier transform of the gap dynamics with $1/(k_Fa_i) = -0.4$ and the BdG eigenenergies $E_k$.}
		\label{fig:vanish}
	  \end{figure}
	  
	  Figure~\ref{fig:vanish}(a) shows the dynamics of the modulus of the spatially averaged order parameter in a system with $r=50$ after quenches with $1/(k_Fa_i) = -0.4 $ (blue solid curve) and $1/(k_Fa_i) = -0.8$ (red dash-dotted curve). After the stronger quench (i.e., $1/(k_Fa_i) = -0.4$, being identical to Fig.~\ref{fig:spatiotemporal}) a decay of the modulus of the order parameter to a value which is much smaller than the ground-state value of the final system $\overline{\Delta}_f$ (solid green line) is seen. Following this decay a plateau evolves where the averaged value of the height of the plateau is labeled by $\Delta_\infty$. Although the height of the plateau is very small compared to $\overline{\Delta}_f$ there are still some oscillations with the transverse confinement frequency $\omega_\perp$ visible. Since we cannot expect an exact vanishing of our numerical solution of a nondissipative model, we will, nevertheless, refer to such a behavior as dynamical vanishing of the gap. A technical discussion of the classifications of the numerical solutions is provided in the Appendix. Overall, our definition of the dynamical vanishing of the gap is in close analogy to what has been labeled  phase I in a homogeneous system \cite{Yuzbashyan2015Quantum}. 
	  
	  Additionally, in contrast to the homogeneous system, there are pronounced spikes visible at $\tau_\text{trap}$. These result from the rephasing of the oscillators, i.e., the quasiparticle occupations and coherences, with equidistant energies spaced by $\delta E = 2 h f_\parallel$ that contribute to the dynamics. Therefore, the time is given by $\tau_\text{trap} = h/\delta E = 1/(2 f_\parallel)$ as has been discussed for the dynamics dominated by a dephasing of these oscillators, also labeled by  phase II \cite{Hannibal2015Quench, Yuzbashyan2015Quantum}.
	  
	  In comparison to the spatiotemporal dynamics in Fig.~\ref{fig:spatiotemporal} we find that $\overline{\Delta}(t)$ precisely maps the rapid decay and the revival of the order parameter. As expected the oscillations towards the center of the trap are not visible; instead we observe an oscillation with the transverse confinement frequency. In general, this oscillation is also visible in $\Delta(\vec{r})$. However, it is not visible in Fig.~\ref{fig:spatiotemporal} due to the temporal spacing of frames since $f_\perp \gg f_\parallel$. Thus, we establish that $\overline{\Delta}(t)$ is suitable to analyze both the initial decay and the evolution of a plateau in the dynamics of the order parameter in a cigar-shaped trap.
	  
	  The weaker quench, i.e., $1/(k_Fa_i) = -0.8$, belongs to phase II and shows an oscillation around a non-vanishing average, slightly smaller than $\overline{\Delta}_f$. The higher-frequency parts are given by the transverse trapping energy $\hbar \omega_\perp$. For an exhaustive discussion of phase II in an inhomogeneous system we refer the reader to our previous publication \cite{Hannibal2015Quench}. In addition, comparing the two quenches we see that a larger quench leads to a faster initial decay of the order parameter which we will come back to later.
	  
	  In order to gain further insight into the dynamics we show the Fourier transform of the dynamically vanishing amplitude of the order parameter in Fig. \ref{fig:vanish}(b). We observe a spectrum with equidistant Fourier components spaced by $2f_\parallel$ which decay with increasing energy. The equidistant peaks evolve as a result of a nonlinear behavior of the equations of motion. While for a small quench, i.e., in the linear phase II, the Fourier components are given by the BdG quasiparticle energies (green dots) the frequencies for a large quench are given by the difference between the energies of the quasiparticle states contributing to the dynamics. To be precise, each occupation and coherence oscillates dominantly with a frequency given by twice the difference between the corresponding single-particle energy and the chemical potential $\mu$. Since the final system is located in the BCS regime, i.e., $\mu \approx E_F$, this difference is always a multiple of $\hbar \omega_\parallel$. Overall, we obtain in phase I a spectrum with equidistant peaks dominated by non-linear effects. We refer to such a spectrum as ``quasi-ungapped'' spectrum, since the gap of the spectrum is given by the energy spacing of the equidistant peaks which is much smaller than the quasiparticle gap of the ground state. 
	  
	  In such a case of an equidistant quasi-ungapped spectrum a Fourier series directly connects the initial decay in the time domain with the width of the distribution of the frequency components. We label this width by $\beta$ and determine it by an exponential fit in the time domain, which we discuss in the Appendix. In frequency space a physical understanding of the width of the decay can be obtained when considering the BCS pairing properties: Cooper pairing of atoms takes place in an interval around the chemical potential determined by the BdG wave functions. From the solution of the BdG equation we see that the width of this interval is determined by the gap $\Delta_k$ of state $k$ at the chemical potential $\mu$. Further, the pairing strength continuously decreases in the given interval with increasing distance to the chemical potential due to the form of the BdG wave functions. This implies that the amplitude of the oscillation of the occupations and coherences decays with increasing distance to the chemical potential. Therefore, considering Eq.~\eqref{eq:Delta_Bogolon} also the Fourier coefficients of the dynamical vanishing gap decay with increasing energy over a width determined by $\Delta_k$.
	  
	  Overall, we find that the width $\beta$ is connected to the pairing properties at the chemical potential which can be characterized by the smallest BdG eigenenergy $E_\text{min}$, since $\xi_k \approx \mu$ at the chemical potential. Hence, we obtain that $\beta \propto E_\text{min}$ for the dynamical vanishing of phase I. We have confirmed numerically that there is no additional dependence on the aspect ratio, i.e., for a given $E_\text{min}$ $\beta$ is identical for any aspect ratio $r$.
	  
	  Summarizing, from Fig.~\ref{fig:vanish} we have extracted the relevant physical properties to understand the dynamical vanishing in an inhomogeneous system. We have shown that the dynamics has a nonlinear character and we have related the dynamical behavior to the ground-state properties of the final system for a given quench strength and aspect ratio. We have shown that for a given quench the initial decay characterized by $\beta$ is proportional to the pairing properties in the final state, i.e., $E_\text{min}$ which, thus, will be an important variable for the subsequent analysis. With this fundamental understanding we now turn to the transition between the two different ``phases'' and investigate the impact of different aspect ratios.

	  \subsection{Impact of the aspect ratio}\label{sec:aspect_ratio}
	  
	  In the following we will systematically investigate the impact of the aspect ratio of the cloud on the emergence of a dynamical vanishing of the spatially averaged order parameter. We will first establish numerical criteria necessary for the analysis and then evaluate the dynamics for various aspect ratios and again connect the observed behavior to the ground-state pairing properties characterized by $E_\text{min}$.
	  
	  For the subsequent analysis of the dynamical vanishing of the order parameter $\Delta_\infty$ will be the central quantity. We obtain $\Delta_\infty$ from our numerical data by fitting an exponential decay $f(t) = \text{exp}(-\beta t) + \alpha$ in the beginning and by checking whether a flat plateau evolves afterwards. For this we require $\beta$ to be large enough (criterion A) and an oscillatory behavior to be nonvisible in the subsequent dynamics (criterion B).  If a plateau is identified we take $\Delta_\infty$ as the height of this plateau otherwise $\Delta_\infty$ is the arithmetic mean over the whole calculation time. A discussion of details and implications of our numerical procedure is provided in the Appendix.
	  
	  \begin{figure}[t]
		\includegraphics[width=1\columnwidth]{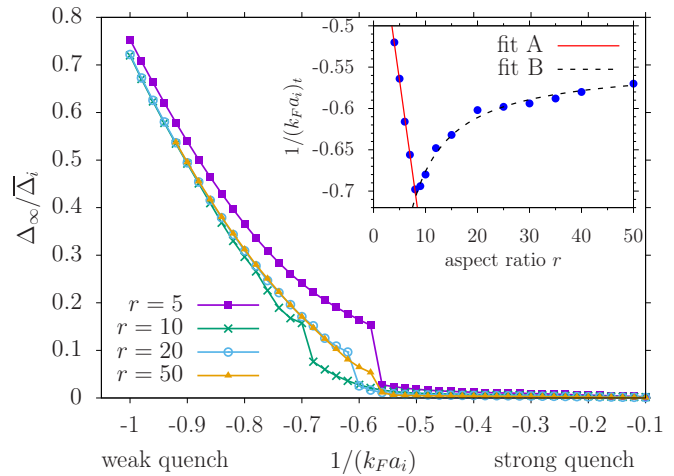}
		\caption{ $\Delta_\infty$ normalized to the ground-state gap of the initial system $\overline{\Delta}_i$ for different quenches with $1/(k_Fa_f) = -1.45$ and aspect ratios $r$. Inset: quench parameter $1/(k_Fa_i)_t$ at the transition to the dynamical vanishing vs. the aspect ratio $r$. The red solid line (fit A) shows a linear fit to data for $r\le 8$ and the black dashed line (fit B) is an inverse proportional fit to data for $r \ge 9$.}
		\label{fig:aspect_ratio}
	  \end{figure}
	  
	  In the following, we carry out an analysis of $\Delta_\infty$ for a range of quenches which all end at the same $1/(k_Fa_f) = -1.45$ where we vary the aspect ratio of the systems and keep the parameters $s=3.5$ and $\omega_\parallel =2\pi\, \cdot 120\,$Hz fixed. In Fig. \ref{fig:aspect_ratio} we show results for four different aspect ratios which represent the effects obtained from our numerical data. For small quenches, i.e., small $1/(k_Fa_i)$, we find a decreasing of $\Delta_\infty$ with increasing quench strength, i.e., larger $1/(k_Fa_i)$, as has been predicted before \cite{Yuzbashyan2015Quantum, Scott2012Rapid}. Increasing the quench strength further, we observe a drop in $\Delta_\infty$ where the position of this visible drop is altered depending on the aspect ratio. The drop implies that the specified conditions for a vanishing order parameter are met and $\Delta_\infty$ is taken as the height of the identified plateau. Furthermore, we see that the height of the obtained plateau continuously tends to zero where it almost coincides for all aspect ratios. In the following, we will investigate the dependence of the drop on the aspect ratio which marks the transition to a vanishing order parameter.
	  
	  For $r=5$ (purple boxes) we find a distinct drop at $1/(k_Fa_i)_t \approx -0.56$ to a plateau with a small height. When increasing the aspect ratio to $r=10$ (green crosses) we observe that the transition to a plateau behavior is shifted to smaller quench strength, i.e., $1/(k_Fa_i)_t \approx -0.69$. At the same time the height of the plateau after the transition is increased. In contrast, increasing the aspect ratio even further to $r=20$ (blue circles) and $r=50$ (orange triangles) leads to a reversal in the shift of the necessary quench strength for the transition. For $r=50$ it is given by $1/(k_Fa_i)_t \approx -0.56$, while the height of the plateau is reduced, since it is determined by the same curve for all aspect ratios which is set by the nonlinear behavior of the equations of motion.
	  
	  In order to investigate the shifts of the transition to a dynamical vanishing in more detail, we plot the initial coupling parameter at the transition point $1/(k_Fa_i)_t$ in dependence of the aspect ratio in the inset of Fig.~\ref{fig:aspect_ratio}. For small aspect ratios $r \le 8$ we find that $1/(k_Fa_i)_t$ decreases linearly which is illustrated by the linear fit A. In contrast, for $r \ge 9$ the transition quench strength increases again and shows an $a/r + b$ (fit B) dependence, where $b \approx -0.55$ is in agreement with the maximum value of $1/(k_Fa_i)_t$ in the main part of Fig.~\ref{fig:aspect_ratio}. It provides a limit of the quench strength necessary for aspect ratios $r \ge 50$. In the following, we will investigate these two confinement-induced shifts of the critical value of the quench strength characterized by $1/(k_Fa_i)$, responsible for the transition to the dynamical vanishing. To this end, we will analyze the dominant energy and, hence, time scales of the system. This includes the dependence of the initial decay constant $\beta$, which turns out to control the behavior according to fit A, as well as the period of the Higgs mode in phase II, which will control the emergence of a dynamical vanishing and the visibility of a plateau in the range of aspect ratios where fit B applies.
	  
\begin{figure}[t]
		\includegraphics[width=1\columnwidth]{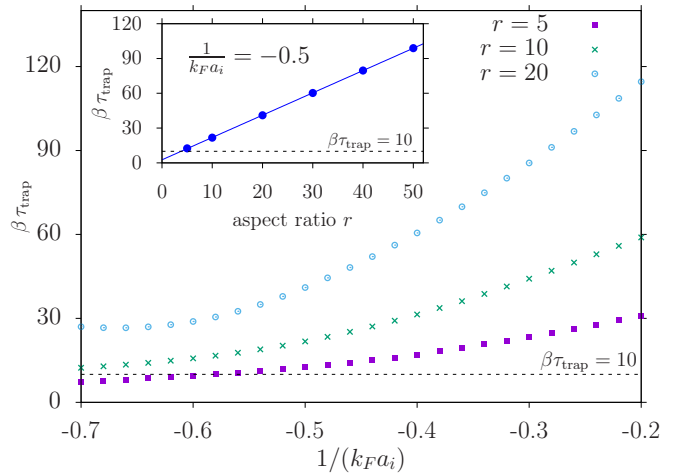}
		\caption{Time scale of the initial decay with respect to the trap time $\beta \tau_\text{trap}$ for various quenches to $1/(k_F a_f) = -1.45$ and $\omega_\parallel = \text{const.}$ for $r=5,10$ and $20$. Inset: $\beta \tau_\text{trap}$ vs. the aspect ratio $r$ at fixed quench conditions; the dashed black line marks the condition $\beta \tau_\text{trap} = 10$ in either figure.}
		\label{fig:scaling_beta}
	  \end{figure}

	  In order to observe a vanishing of the order parameter the initial decay needs to be sufficiently fast compared to the trap time $\tau_\text{trap}$ (cf. criterion A). Therefore, we start by investigating the initial decay characterized by $\beta$ with respect to the trap time $\tau_\text{trap}$ in dependence of the initial coupling $1/(k_Fa_i)$ and the aspect ratio $r$. Figure~\ref{fig:scaling_beta} shows $\beta \tau_\text{trap}$ for a range of different quenches which are identical to those in Fig.~\ref{fig:aspect_ratio}. We find a nonlinear increase of the initial decay rate with an increased quench strength for all aspect ratios. That is, a stronger quench leads to a faster decay as expected for our initial value problem: the initially excited occupations and coherences depend on the order parameter in the initial system. Since we keep the final system identical the initial system for a larger quench is located closer to the BCS-BEC crossover and, hence, features a larger gap. This leads to a wider pairing interval determined by the BdG coefficients $u_m,v_m$ and, therefore, to larger initial values which are distributed over a wider range of energies. This carries over to the spectral properties of the dynamics and thus leads to a larger $\beta$ and, hence, a faster initial decay. Therefore, we establish the criterion $\beta \tau_\text{trap} > 10$ as a necessary condition for the vanishing to occur, as discussed in the Appendix.
	  
	  In the case of a system with an aspect ratio of $r=5$ this criterion is met for $1/(k_Fa_i) > -0.56$. This matches with the onset of the dynamical vanishing as seen in Fig.~\ref{fig:aspect_ratio} and applies to all systems which show the behavior of fit A.  The linear behavior of fit A is inherited from the linear dependence of $\beta \tau_\text{trap}$ on the aspect ratio which we show in the inset of Fig.~\ref{fig:scaling_beta} at a fixed quench strength. Here, the solid blue line is a linear fit to the data points. Since the dependence of $\beta \tau_\text{trap}$ on the quench strength is rather small in the range of the shift of $1/(k_Fa_i)_t$ in Fig.~\ref{fig:aspect_ratio} the dependence of $1/(k_Fa_i)_t$ on $r$ is dominated by the dependence of $\beta \tau_\text{trap}$ on the aspect ratio which results in the observed linear dependence of fit A due to criterion A.
	  
	  In contrast, for the other shown aspect ratios in Fig.~\ref{fig:scaling_beta} ($r=10,20$) the condition $\beta \tau_\text{trap} >10$ is fulfilled for any quench strength with $1/(k_Fa_i) > -0.7$. Nevertheless, in comparison with Fig.~\ref{fig:aspect_ratio} one finds that this does not necessarily lead to a dynamical vanishing of the gap. Furthermore, in the inset of Fig.~\ref{fig:aspect_ratio} we found a different dependence of $1/(k_Fa_i)_t$ on the aspect ratio $r$ for these systems. This suggests that a different effect, which turns out to be connected to criterion B, is responsible for the transition point to the dynamical vanishing. Since these systems are located at the transition between phase I and phase II we go back to a system from phase II [cf. red dash-dotted line in Fig.~\ref{fig:vanish}(a)] and discuss the dependence of the relevant energy scale on the aspect ratio in this phase.
	  
	  If the period of the Higgs mode $T_\text{Higgs}$ is large compared to the trap time, i.e., $T_\text{Higgs} / \tau_\text{trap} \gg 1$, a plateau after the initial decay can become visible (cf. Fig.~\ref{fig:fit} in the Appendix). Additionally to the criterion A of a fast initial decay, characterized by $\beta \tau_\text{trap}$, we require the features of the linear Higgs mode of phase II to be non-visible in the dynamics. This is expressed through the condition $T_\text{Higgs}/ \tau_\text{trap} \gg 1$ (criterion B). If this second condition is not met, we observe an oscillation of the spatially averaged order parameter around the mean value $\Delta_\infty$ even if the initial decay is sufficiently fast.
	  
	  In order to determine the period of the linear Higgs mode we consider the dominant energy scale in phase II which is given by the main frequency of the amplitude (Higgs) mode $f_\text{Higgs}$. Previous studies \cite{Hannibal2015Quench, Yuzbashyan2015Quantum} showed that $f_\text{Higgs}$ depends on the quench strength: in the case of a very small quench $f_\text{Higgs}$ is given by $2E_\text{min}$. Further, for all quenches from ''phase II`` $f_\text{Higgs}$ is closely connected to the mean value $\Delta_\infty$ of the oscillations \footnote{In the homogeneous dynamical BCS theory both the frequency of the oscillations and the average value is given by the gap.}. Therefore, $f_\text{Higgs}$ decreases with increasing excitation strength as can be seen from Fig.~\ref{fig:aspect_ratio}. However, the decrease of $\Delta_\infty$ occurs in the same manner for all aspect ratios and, hence, does not introduce a significant dependence of $f_\text{Higgs}$ on the aspect ratio. As a result, the dependence of $f_\text{Higgs}$ on the aspect ratio is in good approximation solely inherited from the dependence of $E_\text{min}$ on the aspect ratio.
	  
	  We investigate this dependence in the following sections and find that the minimal quasiparticle energy $E_\text{min}$ increases with the aspect ratio. Therefore, the period of the Higgs mode $T_\text{Higgs}$ decreases with increasing aspect ratio and it turns out that this decrease is in good agreement with the $1/r$ behavior of fit B in the inset of Fig.~\ref{fig:aspect_ratio}, which we will come back to in Sec.~\ref{sec:large}. Hence now the transition is determined by the criterion B.
	  
	  
	  Summarizing, we have analyzed the confinement-induced effects on the continuous transition to the dynamical vanishing of the order parameter in a cigar-shaped ultracold Fermi gas. Exploiting our introduced classification, we have formulated two criteria which need to be fulfilled in order to observe a dynamical vanishing characterized by the emergence of a plateau. These two criteria depend contrarily on the aspect ratio and,  hence, we distinguish the following two cases. For small aspect ratios $r \lesssim 8$ the transition is determined by the initial decay constant $\beta \propto E_\text{min}$. This leads to a linear shift of the transition in dependence of the order parameter according to fit A. For larger aspect ratios $r \gtrsim 9$ (fit B) the transition is determined by the period of the Higgs mode where again the relevant dependence is solely given by $E_\text{min}$. Since we have shown that both criteria depend on $E_\text{min}$ we will provide a more general study of this key variable in the next sections.
	  
	  
	  
	  \subsection{Scaling properties}\label{sec:scaling}
	  
	  In this section we will consider the dependence of $E_\text{min}$ on the aspect ratio $r$ in a more general situation where the aspect ratio $r = \omega_\perp / \omega_\parallel$ is changed as $r \rightarrow \lambda r$, where we now keep a fixed particle number of $N=1000$ as opposed to the parameter $s$. In order to achieve this scaling of the aspect ratio $r$ we consider three possible options which are shown in Table~\ref{tab:scaling}. For each option we numerically calculate $E_\text{min}$, $E_F$, and $\beta$ and extract the scaling behavior by fits analog to those in the inset of Fig.~\ref{fig:scaling_beta}. 
	  
	  Option one is to keep $\omega_\parallel$ fixed, which is similar to the previous method but now we keep the particle number $N$ fixed, option two ensures a constant $\omega_\perp$, while option three leaves the volume of the trap unchanged. Strictly speaking, all scaling possibilities could already be deduced from options one and two but we add option three for illustration. Table~\ref{tab:scaling} lists the exponents $p$ of the corresponding dependence of $E_\text{min}\tau_\text{trap} \rightarrow \lambda^p E_\text{min}\tau_\text{trap}$ for each option.
	  
	  \begin{table}[h]
	   \begin{tabular}{l|c|c|c|c||c}
	  way of scaling & $\omega_\parallel$ & $\omega_\perp$ & $\tau_\text{trap}$ & $E_F$ & $E_\text{min} \tau_\text{trap}$  \\ \hline \hline
	     1: $\omega_\parallel = \text{const}$	    & 0 & 1 & 0 & 2/3 & 2/3 \\ \hline
	     2: $\omega_\perp = \text{const}$&-1 & 0 & 1 & -1/3 & 2/3  \\ \hline
	     3: Vol$\,=\,$const & -2/3 & 1/3 & 2/3 &0 & 2/3  \\ 
	     
	   \end{tabular}
	   \caption{\label{tab:scaling} Scaling exponents $p$ of $\lambda$ for the relevant quantities of a cigar-shaped cloud when scaling the aspect ratio as $r\rightarrow \lambda r$ and fixed particle number $N$.}
	  \end{table}
	  
	  Our numerical study shows that the scaling of both $\beta \tau_\text{trap}$ and $E_\text{min}\tau_\text{trap}$ is --independent of the way in which the scaling of the aspect ratio $r$ is performed-- proportional to $\lambda^{2/3}$. The scaling of $E_\text{min}\tau_\text{trap}$ follows directly from the gap equation~\eqref{eq:gap}: we consider $\Delta_k$ to be located at the bottom of a subband, i.e., $\xi_k \approx \mu$, and omit the term introduced for regularization. Then, according to Eq.~\eqref{eq:Ek} $\Delta_k / E_k  \approx 1$ holds true and the effect of the scaling is determined by $V_{kk'} \propto a \, \omega_\perp \omega_\parallel^{1/2}$. Bearing in mind, that we keep $1/(k_Fa)$ fixed which implies a scaling of the scattering length as $a \propto k_F^{-1} \propto E_F^{-1/2} \propto V^{-1/3}$, this yields in all three options $E_\text{min}\tau_\text{trap} \propto \lambda^{2/3}$, as one can deduce from Table~\ref{tab:scaling}. Additionally, this also implies $\beta \tau_\text{trap} \propto \lambda^{2/3}$, since $\beta \propto E_\text{min}$ as argued in Sec.~\ref{sec:vanish}.
	  
	  However, we point out that from the inset of Fig.~\ref{fig:scaling_beta} we find a different scaling of $\beta\tau_\text{trap}$ which can be explained by the fact that the particle number is changed when keeping the parameter $s = \text{const.}$ If we keep $s=\text{const.}$ and scale the aspect ratio as $r \rightarrow \lambda r$ we find from our analysis that $N \rightarrow \lambda N$ and, hence, $E_F \rightarrow \lambda E_F$. However, in order to generalize our previous findings to any scaling of the aspect ratio thus to experimentally more relevant systems in a cigar-shaped trap, we will discuss the behavior for larger particle numbers in the next subsection.
	  
	   \subsection{Large system}\label{sec:large}
	  We will now consider a cloud of $^6$Li atoms in a cigar-shaped trap with larger particle numbers and calculate the smallest BdG quasiparticle energy $E_\text{min}$ of the system. As argued above the relevant time scales of the dynamics are set by this pairing property for a fixed quench. Hence, in this section we extend our systematic analysis to much larger particle numbers up to $N \sim 10^{5}$ by only calculating the ground-state properties of the given system. A full dynamical calculation of these systems is not possible due to numerical constraints. 
	  
	  
	  \begin{figure}[t]
		\includegraphics[width=1\columnwidth]{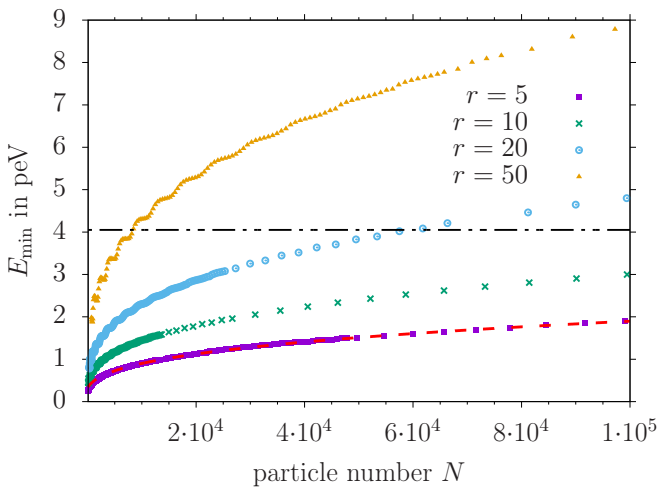}
		\caption{Minimal BdG quasiparticle energy $E_\text{min}$ for various aspect ratios in dependence of the particle number $N$. The red dashed line is a fit $E_\text{min} = a\,N^{1/3} + b$ to the data for $r=5$ and the dash-dotted black line is the value of $E_\text{min}$ for $N=1\cdot 10^6$ obtained from the fit.}
		\label{fig:scaling_N}
	  \end{figure}
	  
	  Figure~\ref{fig:scaling_N} shows the increase of the smallest BdG eigenenergy $E_\text{min}$ in dependence of the particle number $N$, where the aspect ratio is scaled such that $\omega_\parallel =\text{const.}$. The dashed red line shows a fit $E_\text{min} = a \, N^{1/3} + b$ for an aspect ratio $r=5$ and the oscillations visible for small particle numbers are the aforementioned quantum size oscillations \cite{Shanenko2012Atypical}, which can be neglected for the general analysis of the scaling in this paper. Overall, we extract from our numerical data that $E_\text{min} \propto N^{1/3}$ for all aspect ratios $r$ which also implies that $\beta \propto N^{1/3}$.
	  
	  With this result, we now go back to the case discussed in Sec.~\ref{sec:aspect_ratio} where the parameters $s$ and $\omega_\parallel$ are fixed which implies that the particle number scales with $N \propto r$. Combining the general scaling properties of $E_\text{min} \tau_\text{trap} \propto r^{2/3}$ in Table~\ref{tab:scaling} with the dependence on the particle number yields $T_\text{Higgs} \propto 1/f_\text{Higgs} \propto 1/E_\text{min} \propto 1/r$ and $\beta \propto E_\text{min} \propto r$. This is in good agreement with the inset of Fig.~\ref{fig:aspect_ratio} and the inset of Fig.~\ref{fig:scaling_beta}, respectively.	  
	  
	  Furthermore, the black dash-dotted line shows the value of $E_\text{min}$ for $N=1\cdot 10^{6}$ as obtained from the shown fit. From the intersections with the black dash-dotted line obtained in Fig.~\ref{fig:scaling_N} we can choose system parameters with rather small particle numbers which resemble closely the situation in a system with a far larger particle number but smaller aspect ratio.  In the next section we present the calculation of the condensate fraction for a system with $r=50$ and $N = 8428$ for this reason.
	  
	  \subsection{Condensate fraction}\label{sec:condensate}
\begin{figure}[t]
		\includegraphics[width=0.9\columnwidth]{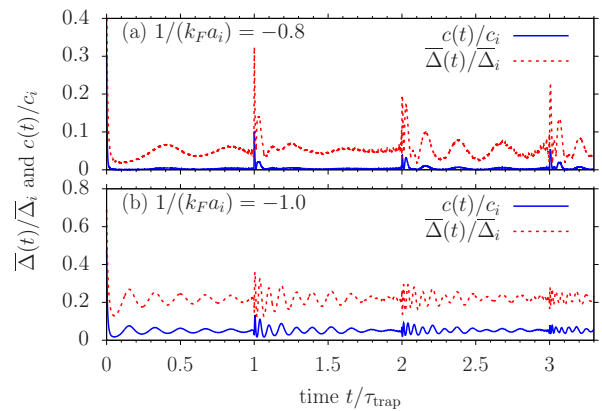}
		\caption{Condensate fraction (blue solid line) and modulus of averaged order parameter (red dashed line) after a quench to $1/(k_Fa_f) = -1.45$ from (a) $1/(k_Fa_i)=-0.8$ and from (b) $1/(k_Fa_i)=-1.0$, with an aspect ratio $r=50$, and a particle number $N = 8428$. The data is normalized to the ground-state value of the initial system $c_i$ and $\overline{\Delta}_i$, respectively.}
		\label{fig:condensate}
	  \end{figure}

	  We obtain the condensate fraction of the system following the usual definition of the condensate fraction $c$ of the order parameter \cite{Leggett2006Quantum}, which yields
	  \begin{equation}\label{eq:condensate}
	   c = \frac{V}{Ng^2} \int d^3 r \left|\Delta(\vec{r})\right|^2,
	  \end{equation}
	  where $V$ is the volume, $N$ the particle number, and $g$ the interaction strength. In Fig.~\ref{fig:condensate} the dynamics of the condensate fraction and the modulus of the averaged gap is shown for a system with an aspect ratio of $r=50$ and a particle number $N =8428$. This is the system that energetically resembles a system with an aspect ratio of $r=5$ and a particle number of $N = 1\cdot 10^6$.
	  	  
	  From Fig.~\ref{fig:condensate}(a) it becomes apparent that the visibility of the signature of the Higgs mode is reduced in the signal of the condensate fraction due to the quadratic dependence on the order parameter. While the signal still contains all frequencies inherited from the Higgs mode, the amplitude of these oscillations is barely visible and the resulting dynamics of the condensate fractions fulfills all requirements to be identified as vanishing.  In the case of a smaller quench, i.e., further away from the transition, the signal of the condensate fraction is large enough to maintain the visibility of the Higgs mode, i.e., its shape and, hence, main frequency, as is shown in Fig.~\ref{fig:condensate}(b). 
	  
	  Overall, the condensate fraction could provide a measure to detect amplitude oscillations of the superfluid gap in an ultracold Fermi gas. However, due to the quadratic dependence on the gap the condensate fraction is already for rather small quenches reduced to $\lesssim 10\%$ of the initial condensate fraction. This can make it challenging to detect oscillations in a time-resolved manner. Further, when analyzing the transition to a vanishing order parameter this will lead to observing a shift of the transition to smaller quench strength. However, we expect the finite-size effects to be qualitatively identical for the condensate fraction as previously discussed for the modulus of the spatially averaged order parameter since all main frequencies are inherited.
	  
	  \section{Conclusion}\label{sec:conclusions}
	  
	  In this paper, we have presented a numerical study of the dynamical vanishing of the order parameter in a cigar-shaped ultracold Fermi gas in the framework of the fully microscopic BdG equations. We have calculated the spatiotemporal dynamics of the order parameter after an interaction quench. We have observed a rapid initial decay as well as a revival which both affect the whole trap. Thus, both can be characterized by the spatially averaged gap. Whereas, after the initial decay we find an oscillation of the order parameter in the longitudinal direction of the harmonic trap which is not seen in the average value due to the continuity equation.
	  
	  Exploiting the spatially averaged order parameter, we have demonstrated that the dynamical vanishing in inhomogeneous systems is robust with respect to the size and aspect ratio of the system while modifications to this vanishing arise from the finite size due to the trap potential. The occurrence of a vanishing order parameter predicted in the homogeneous case is maintained but the precise onset of the vanishing is altered by the pairing properties in the trap. We find that the pairing strength, characterized by the gap at the chemical potential or equivalently by the smallest BdG eigenenergies, increases with the aspect ratio as $r^{2/3} / \tau_\text{trap}$ and with the particle number as $N^{1/3}$.
	  
	  We have found that the transition to a vanishing order parameter in a cigar-shaped ultracold Fermi gas is determined by two criteria which both need to be fulfilled and which depend contrarily on the aspect ratio of the trap. For small aspect ratios the position of the transition is determined by the initial decay rate which needs to be sufficiently fast compared to the characteristic time scale of the trap. In this case we observe that the quench strength necessary for the transition decreases with increasing aspect ratio. However, for smaller quench strengths and larger aspect ratios the signature of the linear Higgs mode from ''phase II`` becomes more prominent which prevents a vanishing to be visible. Then, the transition is determined by the linear Higgs mode which leads to an increasing transition quench strength with increasing aspect ratio.
	  
	  Furthermore, we have shown that the condensate fraction on the BCS side of the BCS-BEC crossover can be a suitable measure to access the Higgs mode in such a cigar-shaped ultracold Fermi gas. Nevertheless, we also find that the condensate fraction will reduce the amplitude of the signal compared to the order parameter and does not mirror the Higgs mode exactly. We expect that this does not introduce new effects but will rather only lead to a shift of the observed transition.
	  
	  \appendix*
	   \section{Fitting procedure}\label{sec:fit}
	   
	    In this Appendix we discuss our fitting routine in order to obtain $\Delta_\infty$ in ''phase I`` and ''phase II`` from our numerical data $\overline{\Delta}(t)$. 
	    
	    \begin{figure}[t]
		\includegraphics[width=0.9\columnwidth]{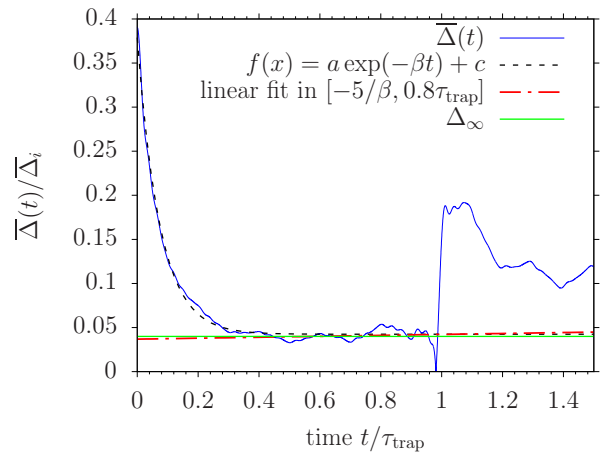}
		\caption{Exemplary illustration of our fitting routines in order to obtain $\Delta_\infty$ for a system with $r=10$ and a quench with $1/(k_Fa_i)=-0.66$. In this system we find a plateau behavior before $\tau_\text{trap}$, which is characterized by the non-vanishing mean value of the plateau.
		\label{fig:fit}}
	    \end{figure}
	    
	    In Fig.~\ref{fig:fit} we show the results of the fitting routines carried out for an exemplary case of $r=10$ and $1/(k_Fa_i) = -0.66$, i.e., a system with a plateau with a non-vanishing height (cf. Fig.~\ref{fig:aspect_ratio}). First, we fit $f(t) =a\, \text{exp}(-\beta t) + c$ to our data (black dotted line) and check whether $\beta\, \tau_\text{trap}/2 > 5$ holds true. If that is the case the initial decay is fast enough that at $t= 0.5\tau_\text{trap}$ the amplitude of the oscillations described by an exponential decay contributes with less than $1\%$ to the total signal and a plateau can become visible before $\tau_\text{trap}$ as shown in Fig.~\ref{fig:fit}.
	    
	    Since the exponential fit is only carried out for the initial decay the offset does not identify the visibility of a plateau correctly in the transition between the two phases. Therefore, if $\beta\, \tau_\text{trap}/2 > 5$ is fulfilled, we carry out a linear fit from $5/\beta$ to $0.8\, \tau_\text{trap}$ and require the relative incline to be less than 10\%. If these criteria are met we find a plateau and choose $\Delta_\infty$ to be the mean value for $5/\beta \le t \le 0.8\,\tau_\text{trap}$; otherwise, we take $\Delta_\infty$ as the arithmetic mean over the whole calculation time. The lower bound of the interval is motivated by the initial exponential decay while the upper bound ensures that the emergence of the peak at $\tau_\text{trap}$ does not hinder a detection of a visible plateau. A linear fit is necessary to rule out oscillatory behavior as, for example, seen in the red dash-dotted line in Fig.~\ref{fig:vanish}. This routine enables us to systematically distinguish between the dynamical vanishing of the gap including the height of the plateau and linear dephasing dynamics of ''phase II``.

	    In mean-field theory a phase transition is obtained in the thermodynamic limit, which has previously been done on the case of the homogeneous system \cite{Yuzbashyan2015Quantum}. In order to investigate the confinement-induced effects to the dynamics in such an ultracold Fermi gas the thermodynamic limit is not suitable. This directly implies that the transition between the ''phase I`` and ''phase II`` takes place continuously in the inhomogeneous system. The above introduced classification quantifies the effects of the two ''phases`` and, hence, enables us to investigate the effects of the aspect ratio to this continuous transition. Although the criteria are well motivated from our observations of the dynamical behavior the exact numerical values for each criteria are within a reasonable range arbitrarily. However, changing the criteria within a reasonable range does not have a qualitative effect on the results presented in this paper. Solely, small quantitative changes such as slightly shifted transition points to the vanishing regime or heights of the plateau occur.
	    

	


\begin{thebibliography}{43}
\expandafter\ifx\csname natexlab\endcsname\relax\def\natexlab#1{#1}\fi
\expandafter\ifx\csname bibnamefont\endcsname\relax
  \def\bibnamefont#1{#1}\fi
\expandafter\ifx\csname bibfnamefont\endcsname\relax
  \def\bibfnamefont#1{#1}\fi
\expandafter\ifx\csname citenamefont\endcsname\relax
  \def\citenamefont#1{#1}\fi
\expandafter\ifx\csname url\endcsname\relax
  \def\url#1{\texttt{#1}}\fi
\expandafter\ifx\csname urlprefix\endcsname\relax\def\urlprefix{URL }\fi
\providecommand{\bibinfo}[2]{#2}
\providecommand{\eprint}[2][]{\url{#2}}

\bibitem[{\citenamefont{Bloch et~al.}(2008)\citenamefont{Bloch, Dalibard, and
  Zwerger}}]{Bloch2008Many}
\bibinfo{author}{\bibfnamefont{I.}~\bibnamefont{Bloch}},
  \bibinfo{author}{\bibfnamefont{J.}~\bibnamefont{Dalibard}}, \bibnamefont{and}
  \bibinfo{author}{\bibfnamefont{W.}~\bibnamefont{Zwerger}},
  \bibinfo{journal}{Rev. Mod. Phys.} \textbf{\bibinfo{volume}{80}},
  \bibinfo{pages}{885} (\bibinfo{year}{2008}).

\bibitem[{\citenamefont{Goldman et~al.}(2016)\citenamefont{Goldman, Budich, and
  Zoller}}]{Goldman2016Topological}
\bibinfo{author}{\bibfnamefont{N.}~\bibnamefont{Goldman}},
  \bibinfo{author}{\bibfnamefont{J.}~\bibnamefont{Budich}}, \bibnamefont{and}
  \bibinfo{author}{\bibfnamefont{P.}~\bibnamefont{Zoller}},
  \bibinfo{journal}{Nature Physics} \textbf{\bibinfo{volume}{12}},
  \bibinfo{pages}{639} (\bibinfo{year}{2016}).

\bibitem[{\citenamefont{Galitski and Spielman}(2013)}]{Galitski2013Spin}
\bibinfo{author}{\bibfnamefont{V.}~\bibnamefont{Galitski}} \bibnamefont{and}
  \bibinfo{author}{\bibfnamefont{I.~B.} \bibnamefont{Spielman}},
  \bibinfo{journal}{Nature} \textbf{\bibinfo{volume}{494}}, \bibinfo{pages}{49}
  (\bibinfo{year}{2013}).

\bibitem[{\citenamefont{Wu et~al.}(2016)\citenamefont{Wu, Zhang, Sun, Xu, Wang,
  Ji, Deng, Chen, Liu, and Pan}}]{Wu2016Realization}
\bibinfo{author}{\bibfnamefont{Z.}~\bibnamefont{Wu}},
  \bibinfo{author}{\bibfnamefont{L.}~\bibnamefont{Zhang}},
  \bibinfo{author}{\bibfnamefont{W.}~\bibnamefont{Sun}},
  \bibinfo{author}{\bibfnamefont{X.-T.} \bibnamefont{Xu}},
  \bibinfo{author}{\bibfnamefont{B.-Z.} \bibnamefont{Wang}},
  \bibinfo{author}{\bibfnamefont{S.-C.} \bibnamefont{Ji}},
  \bibinfo{author}{\bibfnamefont{Y.}~\bibnamefont{Deng}},
  \bibinfo{author}{\bibfnamefont{S.}~\bibnamefont{Chen}},
  \bibinfo{author}{\bibfnamefont{X.-J.} \bibnamefont{Liu}}, \bibnamefont{and}
  \bibinfo{author}{\bibfnamefont{J.-W.} \bibnamefont{Pan}},
  \bibinfo{journal}{Science} \textbf{\bibinfo{volume}{354}},
  \bibinfo{pages}{83} (\bibinfo{year}{2016}).

\bibitem[{\citenamefont{Liao et~al.}(2010)\citenamefont{Liao, Rittner,
  Paprotta, Li, Partridge, Hulet, Baur, and Mueller}}]{Liao2010Spin}
\bibinfo{author}{\bibfnamefont{Y.-a.} \bibnamefont{Liao}},
  \bibinfo{author}{\bibfnamefont{A.~S.~C.} \bibnamefont{Rittner}},
  \bibinfo{author}{\bibfnamefont{T.}~\bibnamefont{Paprotta}},
  \bibinfo{author}{\bibfnamefont{W.}~\bibnamefont{Li}},
  \bibinfo{author}{\bibfnamefont{G.~B.} \bibnamefont{Partridge}},
  \bibinfo{author}{\bibfnamefont{R.~G.} \bibnamefont{Hulet}},
  \bibinfo{author}{\bibfnamefont{S.~K.} \bibnamefont{Baur}}, \bibnamefont{and}
  \bibinfo{author}{\bibfnamefont{E.~J.} \bibnamefont{Mueller}},
  \bibinfo{journal}{Nature} \textbf{\bibinfo{volume}{467}},
  \bibinfo{pages}{567} (\bibinfo{year}{2010}).

\bibitem[{\citenamefont{Giorgini et~al.}(2008)\citenamefont{Giorgini,
  Pitaevskii, and Stringari}}]{Giorgini2008Theory}
\bibinfo{author}{\bibfnamefont{S.}~\bibnamefont{Giorgini}},
  \bibinfo{author}{\bibfnamefont{L.}~\bibnamefont{Pitaevskii}},
  \bibnamefont{and}
  \bibinfo{author}{\bibfnamefont{S.}~\bibnamefont{Stringari}},
  \bibinfo{journal}{Rev. Mod. Phys.} \textbf{\bibinfo{volume}{80}},
  \bibinfo{pages}{1215} (\bibinfo{year}{2008}).

\bibitem[{\citenamefont{Chin et~al.}(2010)\citenamefont{Chin, Grimm, Julienne,
  and Tiesinga}}]{Chin2010Feshbach}
\bibinfo{author}{\bibfnamefont{C.}~\bibnamefont{Chin}},
  \bibinfo{author}{\bibfnamefont{R.}~\bibnamefont{Grimm}},
  \bibinfo{author}{\bibfnamefont{P.}~\bibnamefont{Julienne}}, \bibnamefont{and}
  \bibinfo{author}{\bibfnamefont{E.}~\bibnamefont{Tiesinga}},
  \bibinfo{journal}{Rev. Mod. Phys.} \textbf{\bibinfo{volume}{82}},
  \bibinfo{pages}{1225} (\bibinfo{year}{2010}).

\bibitem[{\citenamefont{Regal et~al.}(2004)\citenamefont{Regal, Greiner, and
  Jin}}]{Regal2004Observation}
\bibinfo{author}{\bibfnamefont{C.~A.} \bibnamefont{Regal}},
  \bibinfo{author}{\bibfnamefont{M.}~\bibnamefont{Greiner}}, \bibnamefont{and}
  \bibinfo{author}{\bibfnamefont{D.~S.} \bibnamefont{Jin}},
  \bibinfo{journal}{Phys. Rev. Lett.} \textbf{\bibinfo{volume}{92}},
  \bibinfo{pages}{040403} (\bibinfo{year}{2004}).

\bibitem[{\citenamefont{Zwerger}(2011)}]{Zwerger2011BCS}
\bibinfo{author}{\bibfnamefont{W.}~\bibnamefont{Zwerger}},
  \emph{\bibinfo{title}{The BCS-BEC crossover and the unitary Fermi gas}}, vol.
  \bibinfo{volume}{836} (\bibinfo{publisher}{Springer}, \bibinfo{year}{2011}).

\bibitem[{\citenamefont{Grimm et~al.}(2000)\citenamefont{Grimm,
  Weidem{\"u}ller, and Ovchinnikov}}]{Grimm2000Optical}
\bibinfo{author}{\bibfnamefont{R.}~\bibnamefont{Grimm}},
  \bibinfo{author}{\bibfnamefont{M.}~\bibnamefont{Weidem{\"u}ller}},
  \bibnamefont{and} \bibinfo{author}{\bibfnamefont{Y.~B.}
  \bibnamefont{Ovchinnikov}}, \bibinfo{journal}{Adv. At., Mol., Opt. Phys.}
  \textbf{\bibinfo{volume}{42}}, \bibinfo{pages}{95} (\bibinfo{year}{2000}).

\bibitem[{\citenamefont{Shanenko et~al.}(2012)\citenamefont{Shanenko, Croitoru,
  Vagov, Axt, Perali, and Peeters}}]{Shanenko2012Atypical}
\bibinfo{author}{\bibfnamefont{A.~A.}~\bibnamefont{Shanenko}},
  \bibinfo{author}{\bibfnamefont{M.~D.} \bibnamefont{Croitoru}},
  \bibinfo{author}{\bibfnamefont{A.~V.}~\bibnamefont{Vagov}},
  \bibinfo{author}{\bibfnamefont{V.~M.}~\bibnamefont{Axt}},
  \bibinfo{author}{\bibfnamefont{A.}~\bibnamefont{Perali}}, \bibnamefont{and}
  \bibinfo{author}{\bibfnamefont{F.~M.}~\bibnamefont{Peeters}},
  \bibinfo{journal}{Phys. Rev. A} \textbf{\bibinfo{volume}{86}},
  \bibinfo{pages}{033612} (\bibinfo{year}{2012}).

\bibitem[{\citenamefont{Polkovnikov et~al.}(2011)\citenamefont{Polkovnikov,
  Sengupta, Silva, and Vengalattore}}]{Polkovnikov2011Colloquium:}
\bibinfo{author}{\bibfnamefont{A.}~\bibnamefont{Polkovnikov}},
  \bibinfo{author}{\bibfnamefont{K.}~\bibnamefont{Sengupta}},
  \bibinfo{author}{\bibfnamefont{A.}~\bibnamefont{Silva}}, \bibnamefont{and}
  \bibinfo{author}{\bibfnamefont{M.}~\bibnamefont{Vengalattore}},
  \bibinfo{journal}{Rev. Mod. Phys.} \textbf{\bibinfo{volume}{83}},
  \bibinfo{pages}{863} (\bibinfo{year}{2011}).

\bibitem[{\citenamefont{Yin and Radzihovsky}(2013)}]{Yin2013Quench}
\bibinfo{author}{\bibfnamefont{X.}~\bibnamefont{Yin}} \bibnamefont{and}
  \bibinfo{author}{\bibfnamefont{L.}~\bibnamefont{Radzihovsky}},
  \bibinfo{journal}{Phys. Rev. A} \textbf{\bibinfo{volume}{88}},
  \bibinfo{pages}{063611} (\bibinfo{year}{2013}).

\bibitem[{\citenamefont{Yin and Radzihovsky}(2016)}]{Yin2016Quench}
\bibinfo{author}{\bibfnamefont{X.}~\bibnamefont{Yin}} \bibnamefont{and}
  \bibinfo{author}{\bibfnamefont{L.}~\bibnamefont{Radzihovsky}},
  \bibinfo{journal}{Phys. Rev. A} \textbf{\bibinfo{volume}{94}},
  \bibinfo{pages}{063637} (\bibinfo{year}{2016}).

\bibitem[{\citenamefont{Collaboration}(2012)}]{Chatrchyan2012Observation}
\bibinfo{author}{\bibfnamefont{C.}~\bibnamefont{Collaboration}},
  \bibinfo{journal}{Physics Letters B} \textbf{\bibinfo{volume}{716}},
  \bibinfo{pages}{30 } (\bibinfo{year}{2012}), ISSN \bibinfo{issn}{0370-2693}.

\bibitem[{\citenamefont{Sooryakumar and Klein}(1980)}]{Sooryakumar1980Raman}
\bibinfo{author}{\bibfnamefont{R.}~\bibnamefont{Sooryakumar}} \bibnamefont{and}
  \bibinfo{author}{\bibfnamefont{M.~V.}~\bibnamefont{Klein}},
  \bibinfo{journal}{Phys. Rev. Lett.} \textbf{\bibinfo{volume}{45}},
  \bibinfo{pages}{660} (\bibinfo{year}{1980}).

\bibitem[{\citenamefont{Littlewood and Varma}(1981)}]{Littlewood1981Gauge}
\bibinfo{author}{\bibfnamefont{P.~B.} \bibnamefont{Littlewood}}
  \bibnamefont{and} \bibinfo{author}{\bibfnamefont{C.~M.} \bibnamefont{Varma}},
  \bibinfo{journal}{Phys. Rev. Lett.} \textbf{\bibinfo{volume}{47}},
  \bibinfo{pages}{811} (\bibinfo{year}{1981}).

\bibitem[{\citenamefont{Endres et~al.}(2012)\citenamefont{Endres, Fukuhara,
  Pekker, Cheneau, Schau{\ss}, Gross, Demler, Kuhr, and
  Bloch}}]{Endres2012Higgs}
\bibinfo{author}{\bibfnamefont{M.}~\bibnamefont{Endres}},
  \bibinfo{author}{\bibfnamefont{T.}~\bibnamefont{Fukuhara}},
  \bibinfo{author}{\bibfnamefont{D.}~\bibnamefont{Pekker}},
  \bibinfo{author}{\bibfnamefont{M.}~\bibnamefont{Cheneau}},
  \bibinfo{author}{\bibfnamefont{P.}~\bibnamefont{Schau{\ss}}},
  \bibinfo{author}{\bibfnamefont{C.}~\bibnamefont{Gross}},
  \bibinfo{author}{\bibfnamefont{E.}~\bibnamefont{Demler}},
  \bibinfo{author}{\bibfnamefont{S.}~\bibnamefont{Kuhr}}, \bibnamefont{and}
  \bibinfo{author}{\bibfnamefont{I.}~\bibnamefont{Bloch}},
  \bibinfo{journal}{Nature} \textbf{\bibinfo{volume}{487}},
  \bibinfo{pages}{454} (\bibinfo{year}{2012}).

\bibitem[{\citenamefont{Papenkort et~al.}(2007)\citenamefont{Papenkort, Axt,
  and Kuhn}}]{Papenkort2007Coherent}
\bibinfo{author}{\bibfnamefont{T.}~\bibnamefont{Papenkort}},
  \bibinfo{author}{\bibfnamefont{V.~M.} \bibnamefont{Axt}}, \bibnamefont{and}
  \bibinfo{author}{\bibfnamefont{T.}~\bibnamefont{Kuhn}},
  \bibinfo{journal}{Phys. Rev. B} \textbf{\bibinfo{volume}{76}},
  \bibinfo{pages}{224522} (\bibinfo{year}{2007}).

\bibitem[{\citenamefont{Papenkort et~al.}(2008)\citenamefont{Papenkort, Kuhn,
  and Axt}}]{Papenkort2008Coherent}
\bibinfo{author}{\bibfnamefont{T.}~\bibnamefont{Papenkort}},
  \bibinfo{author}{\bibfnamefont{T.}~\bibnamefont{Kuhn}}, \bibnamefont{and}
  \bibinfo{author}{\bibfnamefont{V.~M.} \bibnamefont{Axt}},
  \bibinfo{journal}{Phys. Rev. B} \textbf{\bibinfo{volume}{78}},
  \bibinfo{pages}{132505} (\bibinfo{year}{2008}).

\bibitem[{\citenamefont{Matsunaga et~al.}(2014)\citenamefont{Matsunaga, Tsuji,
  Fujita, Sugioka, Makise, Uzawa, Terai, Wang, Aoki, and
  Shimano}}]{Matsunaga2014Light}
\bibinfo{author}{\bibfnamefont{R.}~\bibnamefont{Matsunaga}},
  \bibinfo{author}{\bibfnamefont{N.}~\bibnamefont{Tsuji}},
  \bibinfo{author}{\bibfnamefont{H.}~\bibnamefont{Fujita}},
  \bibinfo{author}{\bibfnamefont{A.}~\bibnamefont{Sugioka}},
  \bibinfo{author}{\bibfnamefont{K.}~\bibnamefont{Makise}},
  \bibinfo{author}{\bibfnamefont{Y.}~\bibnamefont{Uzawa}},
  \bibinfo{author}{\bibfnamefont{H.}~\bibnamefont{Terai}},
  \bibinfo{author}{\bibfnamefont{Z.}~\bibnamefont{Wang}},
  \bibinfo{author}{\bibfnamefont{H.}~\bibnamefont{Aoki}}, \bibnamefont{and}
  \bibinfo{author}{\bibfnamefont{R.}~\bibnamefont{Shimano}},
  \bibinfo{journal}{Science} \textbf{\bibinfo{volume}{345}},
  \bibinfo{pages}{1145} (\bibinfo{year}{2014}).

\bibitem[{\citenamefont{Matsunaga et~al.}(2013)\citenamefont{Matsunaga, Hamada,
  Makise, Uzawa, Terai, Wang, and Shimano}}]{Matsunaga2013Higgsa}
\bibinfo{author}{\bibfnamefont{R.}~\bibnamefont{Matsunaga}},
  \bibinfo{author}{\bibfnamefont{Y.~I.} \bibnamefont{Hamada}},
  \bibinfo{author}{\bibfnamefont{K.}~\bibnamefont{Makise}},
  \bibinfo{author}{\bibfnamefont{Y.}~\bibnamefont{Uzawa}},
  \bibinfo{author}{\bibfnamefont{H.}~\bibnamefont{Terai}},
  \bibinfo{author}{\bibfnamefont{Z.}~\bibnamefont{Wang}}, \bibnamefont{and}
  \bibinfo{author}{\bibfnamefont{R.}~\bibnamefont{Shimano}},
  \bibinfo{journal}{Phys. Rev. Lett.} \textbf{\bibinfo{volume}{111}},
  \bibinfo{pages}{057002} (\bibinfo{year}{2013}).

\bibitem[{\citenamefont{Matsunaga and
  Shimano}(2012)}]{Matsunaga2012Nonequilibrium}
\bibinfo{author}{\bibfnamefont{R.}~\bibnamefont{Matsunaga}} \bibnamefont{and}
  \bibinfo{author}{\bibfnamefont{R.}~\bibnamefont{Shimano}},
  \bibinfo{journal}{Phys. Rev. Lett.} \textbf{\bibinfo{volume}{109}},
  \bibinfo{pages}{187002} (\bibinfo{year}{2012}).

\bibitem[{\citenamefont{Fr\"ohlich et~al.}(2011)\citenamefont{Fr\"ohlich, Feld,
  Vogt, Koschorreck, Zwerger, and K\"ohl}}]{Froehlich2011Radio}
\bibinfo{author}{\bibfnamefont{B.}~\bibnamefont{Fr\"ohlich}},
  \bibinfo{author}{\bibfnamefont{M.}~\bibnamefont{Feld}},
  \bibinfo{author}{\bibfnamefont{E.}~\bibnamefont{Vogt}},
  \bibinfo{author}{\bibfnamefont{M.}~\bibnamefont{Koschorreck}},
  \bibinfo{author}{\bibfnamefont{W.}~\bibnamefont{Zwerger}}, \bibnamefont{and}
  \bibinfo{author}{\bibfnamefont{M.}~\bibnamefont{K\"ohl}},
  \bibinfo{journal}{Phys. Rev. Lett.} \textbf{\bibinfo{volume}{106}},
  \bibinfo{pages}{105301} (\bibinfo{year}{2011}).

\bibitem[{\citenamefont{Clark et~al.}(2015)\citenamefont{Clark, Ha, Xu, and
  Chin}}]{Clark2015Quantum}
\bibinfo{author}{\bibfnamefont{L.~W.} \bibnamefont{Clark}},
  \bibinfo{author}{\bibfnamefont{L.-C.} \bibnamefont{Ha}},
  \bibinfo{author}{\bibfnamefont{C.-Y.} \bibnamefont{Xu}}, \bibnamefont{and}
  \bibinfo{author}{\bibfnamefont{C.}~\bibnamefont{Chin}},
  \bibinfo{journal}{Phys. Rev. Lett.} \textbf{\bibinfo{volume}{115}},
  \bibinfo{pages}{155301} (\bibinfo{year}{2015}).

\bibitem[{\citenamefont{Yuzbashyan et~al.}(2015)\citenamefont{Yuzbashyan,
  Dzero, Gurarie, and Foster}}]{Yuzbashyan2015Quantum}
\bibinfo{author}{\bibfnamefont{E.~A.} \bibnamefont{Yuzbashyan}},
  \bibinfo{author}{\bibfnamefont{M.}~\bibnamefont{Dzero}},
  \bibinfo{author}{\bibfnamefont{V.}~\bibnamefont{Gurarie}}, \bibnamefont{and}
  \bibinfo{author}{\bibfnamefont{M.~S.} \bibnamefont{Foster}},
  \bibinfo{journal}{Phys. Rev. A} \textbf{\bibinfo{volume}{91}},
  \bibinfo{pages}{033628} (\bibinfo{year}{2015}).

\bibitem[{\citenamefont{Yuzbashyan et~al.}(2006)\citenamefont{Yuzbashyan,
  Tsyplyatyev, and Altshuler}}]{Yuzbashyan2006Relaxationa}
\bibinfo{author}{\bibfnamefont{E.~A.} \bibnamefont{Yuzbashyan}},
  \bibinfo{author}{\bibfnamefont{O.}~\bibnamefont{Tsyplyatyev}},
  \bibnamefont{and} \bibinfo{author}{\bibfnamefont{B.~L.}
  \bibnamefont{Altshuler}}, \bibinfo{journal}{Phys. Rev. Lett.}
  \textbf{\bibinfo{volume}{96}}, \bibinfo{pages}{097005}
  (\bibinfo{year}{2006}).

\bibitem[{\citenamefont{Yuzbashyan and Dzero}(2006)}]{Yuzbashyan2006Dynamical}
\bibinfo{author}{\bibfnamefont{E.~A.}~\bibnamefont{Yuzbashyan}} \bibnamefont{and}
  \bibinfo{author}{\bibfnamefont{M.}~\bibnamefont{Dzero}},
  \bibinfo{journal}{Phys. Rev. Lett.} \textbf{\bibinfo{volume}{96}},
  \bibinfo{pages}{230404} (\bibinfo{year}{2006}).

\bibitem[{\citenamefont{Yuzbashyan et~al.}(2005)\citenamefont{Yuzbashyan,
  Altshuler, Kuznetsov, and Enolskii}}]{Yuzbashyan2005Nonequilibrium}
\bibinfo{author}{\bibfnamefont{E.~A.} \bibnamefont{Yuzbashyan}},
  \bibinfo{author}{\bibfnamefont{B.~L.} \bibnamefont{Altshuler}},
  \bibinfo{author}{\bibfnamefont{V.~B.} \bibnamefont{Kuznetsov}},
  \bibnamefont{and} \bibinfo{author}{\bibfnamefont{V.~Z.}
  \bibnamefont{Enolskii}}, \bibinfo{journal}{Phys. Rev. B}
  \textbf{\bibinfo{volume}{72}}, \bibinfo{pages}{220503}
  (\bibinfo{year}{2005}).

\bibitem[{\citenamefont{Barankov and
  Levitov}(2006)}]{Barankov2006Synchronization}
\bibinfo{author}{\bibfnamefont{R.~A.} \bibnamefont{Barankov}} \bibnamefont{and}
  \bibinfo{author}{\bibfnamefont{L.~S.} \bibnamefont{Levitov}},
  \bibinfo{journal}{Phys. Rev. Lett.} \textbf{\bibinfo{volume}{96}},
  \bibinfo{pages}{230403} (\bibinfo{year}{2006}).

\bibitem[{\citenamefont{Bruun et~al.}(1999)\citenamefont{Bruun, Castin, Dum,
  and Burnett}}]{Bruun1999BCS}
\bibinfo{author}{\bibfnamefont{G.}~\bibnamefont{Bruun}},
  \bibinfo{author}{\bibfnamefont{Y.}~\bibnamefont{Castin}},
  \bibinfo{author}{\bibfnamefont{R.}~\bibnamefont{Dum}}, \bibnamefont{and}
  \bibinfo{author}{\bibfnamefont{K.}~\bibnamefont{Burnett}},
  \bibinfo{journal}{Eur. Phys. J. D } \textbf{\bibinfo{volume}{7}} \bibinfo{pages}{433}
  (\bibinfo{year}{1999}).

\bibitem[{\citenamefont{Bruun}(2014)}]{Bruun2014Long}
\bibinfo{author}{\bibfnamefont{G.}~\bibnamefont{Bruun}},
  \bibinfo{journal}{Phys. Rev. A} \textbf{\bibinfo{volume}{90}},
  \bibinfo{pages}{023621} (\bibinfo{year}{2014}).

\bibitem[{\citenamefont{Scott et~al.}(2012)\citenamefont{Scott, Dalfovo,
  Pitaevskii, and Stringari}}]{Scott2012Rapid}
\bibinfo{author}{\bibfnamefont{R.~G.}~\bibnamefont{Scott}},
  \bibinfo{author}{\bibfnamefont{F.}~\bibnamefont{Dalfovo}},
  \bibinfo{author}{\bibfnamefont{L.~P.}~\bibnamefont{Pitaevskii}},
  \bibnamefont{and}
  \bibinfo{author}{\bibfnamefont{S.}~\bibnamefont{Stringari}},
  \bibinfo{journal}{Phys. Rev. A} \textbf{\bibinfo{volume}{86}},
  \bibinfo{pages}{053604} (\bibinfo{year}{2012}).

\bibitem[{\citenamefont{Hannibal et~al.}(2015)\citenamefont{Hannibal, Kettmann,
  Croitoru, Vagov, Axt, and Kuhn}}]{Hannibal2015Quench}
\bibinfo{author}{\bibfnamefont{S.}~\bibnamefont{Hannibal}},
  \bibinfo{author}{\bibfnamefont{P.}~\bibnamefont{Kettmann}},
  \bibinfo{author}{\bibfnamefont{M.~D.} \bibnamefont{Croitoru}},
  \bibinfo{author}{\bibfnamefont{A.}~\bibnamefont{Vagov}},
  \bibinfo{author}{\bibfnamefont{V.~M.} \bibnamefont{Axt}}, \bibnamefont{and}
  \bibinfo{author}{\bibfnamefont{T.}~\bibnamefont{Kuhn}},
  \bibinfo{journal}{Phys. Rev. A} \textbf{\bibinfo{volume}{91}},
  \bibinfo{pages}{043630} (\bibinfo{year}{2015}).

\bibitem[{\citenamefont{Zachmann et~al.}(2013)\citenamefont{Zachmann, Croitoru,
  Vagov, Axt, Papenkort, and Kuhn}}]{Zachmann2013Ultrafast}
\bibinfo{author}{\bibfnamefont{M.}~\bibnamefont{Zachmann}},
  \bibinfo{author}{\bibfnamefont{M.~D.} \bibnamefont{Croitoru}},
  \bibinfo{author}{\bibfnamefont{A.}~\bibnamefont{Vagov}},
  \bibinfo{author}{\bibfnamefont{V.~M.} \bibnamefont{Axt}},
  \bibinfo{author}{\bibfnamefont{T.}~\bibnamefont{Papenkort}},
  \bibnamefont{and} \bibinfo{author}{\bibfnamefont{T.}~\bibnamefont{Kuhn}},
  \bibinfo{journal}{New J. Phys.} \textbf{\bibinfo{volume}{15}},
  \bibinfo{pages}{055016} (\bibinfo{year}{2013}).

\bibitem[{\citenamefont{Kettmann et~al.}(2017)\citenamefont{Kettmann, Hannibal,
  Croitoru, Vagov, Axt, and Kuhn}}]{Kettmann2017Spectral}
\bibinfo{author}{\bibfnamefont{P.}~\bibnamefont{Kettmann}},
  \bibinfo{author}{\bibfnamefont{S.}~\bibnamefont{Hannibal}},
  \bibinfo{author}{\bibfnamefont{M.}~\bibnamefont{Croitoru}},
  \bibinfo{author}{\bibfnamefont{A.}~\bibnamefont{Vagov}},
  \bibinfo{author}{\bibfnamefont{V.}~\bibnamefont{Axt}}, \bibnamefont{and}
  \bibinfo{author}{\bibfnamefont{T.}~\bibnamefont{Kuhn}},
  \bibinfo{journal}{Physica C} \textbf{\bibinfo{volume}{533}},
  \bibinfo{pages}{133} (\bibinfo{year}{2017}).

\bibitem[{\citenamefont{Papenkort et~al.}(2009)\citenamefont{Papenkort, Kuhn,
  and Axt}}]{Papenkort2009Nonequilibrium}
\bibinfo{author}{\bibfnamefont{T.}~\bibnamefont{Papenkort}},
  \bibinfo{author}{\bibfnamefont{T.}~\bibnamefont{Kuhn}}, \bibnamefont{and}
  \bibinfo{author}{\bibfnamefont{V.~M.} \bibnamefont{Axt}},
  \bibinfo{journal}{Journal of Physics: Conference Series}
  \textbf{\bibinfo{volume}{193}}, \bibinfo{pages}{012050}
  (\bibinfo{year}{2009}).

\bibitem[{\citenamefont{Chou et~al.}(2017)\citenamefont{Chou, Liao, and
  Foster}}]{Chou2017Twisting}
\bibinfo{author}{\bibfnamefont{Y.-Z.} \bibnamefont{Chou}},
  \bibinfo{author}{\bibfnamefont{Y.}~\bibnamefont{Liao}}, \bibnamefont{and}
  \bibinfo{author}{\bibfnamefont{M.~S.} \bibnamefont{Foster}},
  \bibinfo{journal}{Phys. Rev. B} \textbf{\bibinfo{volume}{95}},
  \bibinfo{pages}{104507} (\bibinfo{year}{2017}).

\bibitem[{\citenamefont{Datta and Bagwell}(1999)}]{Datta1999Can}
\bibinfo{author}{\bibfnamefont{S.}~\bibnamefont{Datta}} \bibnamefont{and}
  \bibinfo{author}{\bibfnamefont{P.~F.} \bibnamefont{Bagwell}},
  \bibinfo{journal}{Superlattices and Microstructures}
  \textbf{\bibinfo{volume}{25}}, \bibinfo{pages}{1233} (\bibinfo{year}{1999}).

\bibitem[{\citenamefont{De~Gennes}(1989)}]{DeGennes1989Superconductivity}
\bibinfo{author}{\bibfnamefont{P.}~\bibnamefont{De~Gennes}},
  \emph{\bibinfo{title}{Superconductivity of metals and alloys}}
  (\bibinfo{publisher}{Addison-Wesley New York}, \bibinfo{year}{1989}).

\bibitem[{\citenamefont{Anderson}(1959)}]{Anderson1959Theory}
\bibinfo{author}{\bibfnamefont{P.~W.} \bibnamefont{Anderson}},
  \bibinfo{journal}{J. Phys. Chem. Solids} \textbf{\bibinfo{volume}{11}},
  \bibinfo{pages}{26} (\bibinfo{year}{1959}).

\bibitem[{\citenamefont{Bruun and Heiselberg}(2002)}]{Bruun2002Cooper}
\bibinfo{author}{\bibfnamefont{G.~M.}~\bibnamefont{Bruun}} \bibnamefont{and}
  \bibinfo{author}{\bibfnamefont{H.}~\bibnamefont{Heiselberg}},
  \bibinfo{journal}{Phys. Rev. A} \textbf{\bibinfo{volume}{65}},
  \bibinfo{pages}{053407} (\bibinfo{year}{2002}).

\bibitem[{\citenamefont{Leggett}(2006)}]{Leggett2006Quantum}
\bibinfo{author}{\bibfnamefont{A.~J.} \bibnamefont{Leggett}},
  \emph{\bibinfo{title}{Quantum liquids: Bose condensation and Cooper pairing
  in condensed-matter systems}} (\bibinfo{publisher}{Oxford University Press},
  \bibinfo{year}{2006}).

\end{thebibliography}

\end{document}